\documentclass[twocolumn,showpacs,aps,prl,superscriptaddress]{revtex4}

\usepackage{xspace}
\usepackage{graphicx}
\usepackage{dcolumn}
\usepackage{amsmath}
\usepackage{epsfig}

\input babarsym.tex

\newcommand{\eett}   {\ensuremath{e^+e^- \to \tautau}\xspace}
\newcommand{\eemm}       {\ensuremath{e^+e^- \to \mumu}\xspace}
\newcommand{\eeqq}       {\ensuremath{e^+e^- \to \qqbar}\xspace}

\newcommand{\tautomu}  {\ensuremath{\tau \to \mu \nunub}\xspace}
\newcommand{\tautoel}  {\ensuremath{\tau \to e \nunub}\xspace}

\newcommand{\BRtauppp}    {\ensuremath{\BR(\tautoppp)}\xspace}
\newcommand{\BRtaukpp}    {\ensuremath{\BR(\tautokpp)}\xspace}
\newcommand{\BRtaukpk}    {\ensuremath{\BR(\tautokpk)}\xspace}
\newcommand{\BRtaukkk}    {\ensuremath{\BR(\tautokkk)}\xspace}
\newcommand{\BRtauphip}    {\ensuremath{\BR(\tautophip)}\xspace}
\newcommand{\BRtauphik}    {\ensuremath{\BR(\tautophik)}\xspace}
\newcommand{\BRtaukkkexphi}    {\ensuremath{\BR(\tautokkk [\mathrm{ex.}\phi])}\xspace}

\newcommand{\roots}        {\ensuremath{\sqrt{s}}\xspace}

\newcommand{\tautohhh} {\ensuremath{\tau^- \to h^-h^-h^+\nu_\tau}\xspace}
\newcommand{\tautoppp} {\ensuremath{\tau^- \to \pi^-\pi^-\pi^+\nu_\tau}\xspace}
\newcommand{\tautokpp} {\ensuremath{\tau^- \to K^-\pi^-\pi^+\nu_\tau}\xspace}
\newcommand{\tautokpk} {\ensuremath{\tau^- \to K^-\pi^-K^+\nu_\tau}\xspace}
\newcommand{\tautokkk} {\ensuremath{\tau^- \to K^-K^-K^+\nu_\tau}\xspace}
\newcommand{\tautophip} {\ensuremath{\tau^- \to \phip \nu_\tau}\xspace}
\newcommand{\tautophippiz} {\ensuremath{\tau^- \to \phip n \pi^0 \nu_\tau}\xspace}

\newcommand{\tautophik} {\ensuremath{\tau^- \to \phik   \nu_\tau}\xspace}

\newcommand{\phip} {\ensuremath{\phi\pi^-}\xspace}
\newcommand{\phik} {\ensuremath{\phi K^-}\xspace}

\newcommand{\Ndatappp} {\ensuremath{(1.5953\pm0.0013) \times 10^6}\xspace}
\newcommand{\Ndatakpp} {\ensuremath{(6.956\pm0.0026) \times 10^4}\xspace}
\newcommand{\Ndatakpk} {\ensuremath{(1.819\pm0.013)  \times 10^4}\xspace}
\newcommand{\Ndatakkk} {\ensuremath{275\pm 17}\xspace}

\newcommand{\Nbkgppp} {\ensuremath{(0.0642\pm 0.0002) \times 10^6}\xspace}
\newcommand{\Nbkgkpp} {\ensuremath{(0.2263\pm 0.0064) \times 10^4}\xspace}
\newcommand{\Nbkgkpk} {\ensuremath{(0.0145\pm 0.0008) \times 10^4}\xspace}
\newcommand{\Nbkgkkk} {\ensuremath{2.5\pm 1.5}\xspace}

\newcommand{\Nsigppp} {\ensuremath{(55.59\pm0.05)\times 10^6}\xspace}
\newcommand{\Nsigkpp} {\ensuremath{(171.5\pm 1.2)\times 10^4}\xspace}
\newcommand{\Nsigkpk} {\ensuremath{(84.71\pm0.66)\times 10^4}\xspace}
\newcommand{\Nsigkkk} {\ensuremath{(9.93\pm 0.84)\times 10^3}\xspace}

\newcommand{\effppp}   {\ensuremath{0.028}\xspace}
\newcommand{\effkpp}   {\ensuremath{0.031}\xspace}
\newcommand{\effkpk}   {\ensuremath{0.035}\xspace}
\newcommand{\effkkk}   {\ensuremath{0.039}\xspace}

\newcommand{\BRppp}   {\ensuremath{ (8.83\pm 0.01 \pm 0.13)\% }\xspace}
\newcommand{\BRkpp}   {\ensuremath{ (0.273\pm 0.002 \pm 0.009)\% }\xspace}
\newcommand{\BRkpk}   {\ensuremath{ (0.1346\pm 0.0010 \pm 0.0036)\% }\xspace}
\newcommand{\BRkkk}   {\ensuremath{ (1.58\pm 0.13 \pm 0.12)\times 10^{-5} }\xspace}

\newcommand{\BRpppPDGa}   {\ensuremath{ (9.13\pm 0.05\pm 0.46)\% }\xspace}
\newcommand{\BRkppPDGa}   {\ensuremath{ (0.33\pm 0.05)\% }\xspace}
\newcommand{\BRkpkPDGa}   {\ensuremath{ (0.154\pm 0.009)\% }\xspace}
\newcommand{\BRkkkPDGa}   {\ensuremath{ < 3.7\times 10^{-5} @90\% CL}\xspace}

\newcommand{\BRphip}   {\ensuremath{ (3.42\pm 0.55 \pm 0.25)\times 10^{-5} }\xspace}
\newcommand{\BRphik}   {\ensuremath{ (3.39\pm 0.20 \pm 0.28)\times 10^{-5} }\xspace}
\newcommand{\BRkkkexphi} {\ensuremath{ < 2.5 \times 10^{-6}} \xspace}

\newcommand{\ntaupair}         {\ensuremath{3.16\times 10^8}\xspace}

\newcommand{\lumi}             {\ensuremath{342\invfb}\xspace}

\def\kk       {\mbox{\tt KK2f}\xspace}
\def\tauola     {\mbox{\tt Tauola}\xspace}

\newcommand{\gevccgevcc}{\ensuremath{{\mathrm{\,Ge\kern -0.1em V^2\!/}c^4}}\xspace}

\newcommand{\evcc}{\ensuremath{{\mathrm{\,e\kern -0.1em V\!/}c^2}}\xspace}
\newcommand{\CM} {\mbox{c.m.}\xspace}

\newcommand{\BABARPubYear}     {06}
\newcommand{\BABARPubNumber}  {060}
\newcommand{\SLACPubNumber} {12686}
\newcommand{\LANLNumber}  {0707.2981 [hep-ex]}

\def\figurebox#1#2#3{%
    \def\arg{#3}%
    \ifx\arg\empty
    {\hfill\vbox{\hsize#2\hrule\hbox to #2{\vrule\hfill\vbox to #1{\hsize#2\vfill}\vrule}\hrule}\hfill}%
    \else
    {\hfill\epsfbox{#3}\hfill}%
    \fi}

\begin{document}

\preprint{\babar-PUB-\BABARPubYear/\BABARPubNumber} 
\preprint{SLAC-PUB-\SLACPubNumber} 
\preprint{\LANLNumber}    

\begin{flushleft}
\babar-PUB-\BABARPubYear/\BABARPubNumber\\
SLAC-PUB-\SLACPubNumber\\
arXiv:\LANLNumber\\
\end{flushleft}

\title{
{\large \bf \boldmath
Exclusive branching fraction measurements of semileptonic tau decays into three charged hadrons,
$\tau^- \rightarrow \phi \pi^- \nu_\tau$ and $\tau^- \rightarrow \phi K^- \nu_\tau$
}}

%
\author{B.~Aubert}
\author{M.~Bona}
\author{D.~Boutigny}
\author{F.~Couderc}
\author{Y.~Karyotakis}
\author{J.~P.~Lees}
\author{V.~Poireau}
\author{V.~Tisserand}
\author{A.~Zghiche}
\affiliation{Laboratoire de Physique des Particules, IN2P3/CNRS et Universit\'e de Savoie, F-74941 Annecy-Le-Vieux, France }
\author{E.~Grauges}
\affiliation{Universitat de Barcelona, Facultat de Fisica, Departament ECM, E-08028 Barcelona, Spain }
% ...
\collaboration{The \babar\ Collaboration}
\noaffiliation

\date{\today}

\begin{abstract}
Using a data sample corresponding to an integrated luminosity of \lumi collected
with the \babar\ detector at the SLAC PEP-II electron-positron storage ring operating at a
center-of-mass energy near 10.58\gev, we measure
$\BRtauppp = \BRppp$,
$\BRtaukpp = \BRkpp$,
$\BRtaukpk = \BRkpk$ and
$\BRtaukkk = \BRkkk$,
where the uncertainties are statistical and systematic, respectively.
Events where the $\pi^+\pi^-$ pair is consistent with coming from a $K^0_S$ are excluded.
These include significant improvements over previous measurements and a first measurement of $\BRtaukkk$ 
in which no resonance structure is assumed.
We also report a first measurement of $\BRtauphip = \BRphip$, a new measurement of $\BRtauphik = \BRphik$
and a first upper limit on $\BRtaukkkexphi$.
\end{abstract}


\maketitle


The weak interaction coupling strength 
between the first and second quark generations~\cite{Cabibbo:1963yz} can be probed with
unprecendented precision in hadronic \mtau decays having net strangeness of unity in the final state
using \eett data collected at the $e^+e^-$ B-factories. 
Inclusive measurements of the strange spectral function, obtained from 
\mtau lepton decays to final states containing kaons,
provide a direct determination of the strange quark mass and 
Cabbibo-Kobayashi-Maskawa (CKM) mixing element $|V_{us}|$~\cite{Gamiz:2004ar}. 
A significant improvement on the precision of  \BRtaukpp in particular,
where one measurement~\cite{Barate:1997ma} disagrees by more than two standard deviations from the
others~\cite{Briere:2003fr,Abbiendi:2004xa},
will have the most immediate impact on 
the determination of these two fundamental Standard Model (SM) parameters using $\tau$ decays~\cite{Maltman:2006ic}.
Measurements of  $\tautophip$ and $\tautophik$ provide an interesting laboratory for studying Okubo-Zweig-Iizuka (OZI)
suppression~\cite{LopezCastro:1996xh}.

Significant improvements on measurements of $\BRtauppp$,
$\BRtaukpp$ and $\BRtaukpk$ are reported together with a first measurement of
$\BRtaukkk$ without resonance assumptions (charge conjugate decays are implied). The data sample
corresponds to an integrated luminosity of $\mathcal{L}$ = \lumi recorded at a center-of-mass (\CM) energy (\roots) near 10.58\gev
using the \babar\ detector at the SLAC \pep2 asymmetric-energy \epem storage ring.
With a luminosity-weighted average cross-section of $\sigma_{\eett}=(0.919\pm0.003)$ nb~\cite{kk,Banerjee:2007is}, 
this corresponds to the production of \ntaupair $\tau$-pair events.

The \babar\ detector~\cite{detector} has a silicon vertex tracker (SVT), drift chamber (DCH),
ring-imaging Cherenkov detector (DIRC) and  electromagnetic calorimeter (EMC) all contained in a 1.5-T solenoid.
The iron flux return of the solenoid is instrumented to identify muons.

\mtau-pair events are simulated with higher-order radiative corrections using the \kk Monte Carlo (MC) generator~\cite{kk}
with $\tau$ decays  simulated with \tauola~\cite{tauola,photos} using measured rates~\cite{PDG}.
The detector response is simulated with \mbox{\tt GEANT4}~\cite{geant}. 
Simulated events for signal as well as SM background
processes~\cite{kk,tauola,photos,Lange:2001uf,Sjostrand:1995iq}
are reconstructed in the same manner as data.
The MC samples are used for selection optimization and systematic error studies.
The number of simulated non-signal events is comparable to the number expected in the data,
with the exception of Bhabha and two-photon events, which are not simulated. Data studies
demonstrate that these backgrounds are negligible.

The basic analysis strategy is to select a pure sample of \tautohhh decays from
\eett events by requiring the partner $\tau^+$ to decay leptonically. Within this sample,
each of the $h^{\pm}$ mesons is uniquely identified as a charged pion or kaon and the decay
categorized as \tautoppp, \tautokpp, \tautokpk or \tautokkk.  An efficiency 
migration matrix, $\mathbf{\cal E}_{ij}$, initially obtained from MC simulations,  is used
to correct for  efficiency losses from all stages of event selection and includes cross-feed between
the four signal channels where $i$ ($j$) is the selected (true) decay mode index.
The $\mathbf{\cal E}_{ij}$ matrix is modified using data control samples of kaons
and pions from $D^{*+}\ra \pi^+D^0,D^0\ra\pi^+K^-$ decays to account for
 small differences between MC and data. The number of decay mode $j$ signal events measured in the sample, $\mathbf{N}^{Sig}_j$, is then:
\begin{eqnarray}
\mathbf{N}_{j}^{Sig} &=& \sum_{i} (\mathbf{\cal E}^{-1})_{ji}\left( \mathbf{N}_{i}^{Data}-\mathbf{N}_{i}^{Bkg}\right) \label{matrix}
\label{eqn1}
\end{eqnarray}
\noindent where $\mathbf{N}^{Data}_i$ is the number of data events selected in decay channel $i$ and
$\mathbf{N}^{Bkg}_i$ is the estimated number of  background events in decay channel $i$ arising from 
sources other than \tautohhh. The branching fraction for decay mode $j$ is then
$\BR_{j} =  \frac{\mathbf{N}_{j}^{Sig}}{2\mathcal{L}\sigma_{\eett}}$.

The \tautohhh sample is obtained by selecting events with four well reconstructed tracks having zero total charge,
where none of the tracks originate from the conversion of photons in the material of the detector.
All four tracks are required to lie within the geometrical acceptance of the EMC and DIRC and, to ensure the
tracks can reach the DIRC, are required to have a transverse momentum of at least 250\mevc.
If any two photons in the event are identified as coming from $\pi^0\ra \gamma\gamma$, the event is
removed from the sample. The event is divided into hemispheres in the \CM by a plane perpendicular
to the thrust axis~\cite{thrust}.
One of the hemispheres, the `lepton-side', is required to contain a  \tautoel or \tautomu decay.
If the hemisphere with an electron (muon) has any neutral particles with more than 1.0 (0.5)\gev 
the event is not selected.
The \eeqq backgrounds  are further reduced to $\sim 0.1\%$ of the 
\tautohhh sample by requiring a thrust magnitude $>0.85$.

Backgrounds from Bhabha and \eemm events with photon conversions are suppressed by requiring
the momentum of the lepton-side track to be less than 80\% of \roots/2
and a requirement that none of the three signal-side tracks pass an electron identification
algorithm. Remaining non-$\tau$ background events, including those from 
two-photon processes, are reduced by requiring the event
 missing \CM transverse momentum to be $>0.009\roots$.

The remaining background in the sample is predominantly from other $\tau$ decays
containing $\pi^0$ and $K^0_S$ mesons.
Events containing a $K_{S}^{0}$ are identified and removed. 
Residual backgrounds from decays having a $\pi^0$ are reduced by requiring
the total  energy in the signal-side hemisphere deposited in the EMC which is 
unassociated with the three charged hadron tracks to be $<$200\mev. 
With these requirements, contributions from 
$\tau^{-} \rightarrow h^{-}n\pi^{0}\nu$  and
$\tau^{-} \rightarrow h^{-}h^{-}h^{+}2\pi^{0}\nu$ are negligible. 

A track in the \tautohhh sample is classified as a kaon using a likelihood approach to combine information from the DIRC,
DCH and SVT with a characteristic kaon identification efficiency of $\sim 80\%$ and pion misidentification rate of $\sim 1\%$.
If the track fails to be identified as a kaon, it is classified as a pion.
Events are selected if 
the signal-side decays are identified as \tautoppp, \tautokpp, \tautokpk or \tautokkk
with decays having a wrong charge combination removed.

The diagonal elements of $\mathbf{\cal E}_{ij}$ excluding particle identification ($\epsilon$),
 numbers of  selected,  background and signal events determined using Eq.~\ref{eqn1}
 are shown in  Table~\ref{table1} for all modes. The increase in $\epsilon$ with number of kaons
 in the decay is associated primarily with the transverse momentum requirement
 and the $K^0_S$ veto.
 The background fraction from $\tau$ decays to $\pi^{-}\pi^{-}\pi^{+}\pi^{0}\nu$
 ($K^{-}\pi^{-}\pi^{+}\pi^{0}\nu$, $K^{-}\pi^{-}K^{+}\pi^{0}\nu$,
 $K^{-}K^{-}K^{+}\pi^{0}\nu$)  in the $\pi^{-}\pi^{-}\pi^{+}$
 ($K^{-}\pi^{-}\pi^{+}$, $K^{-}\pi^{-}K^{+}$, $K^{-}K^{-}K^{+}$)
 candidate sample is estimated to be ($3.6\pm0.3$)\% ($(2.3\pm0.4)\%$, $(0.4\pm0.1)\%$,$<5.0\%$).  
Non-$\tau$ backgrounds comprise less than 0.5\% of each channel's final event sample.

 The component of  $\mathbf{\cal E}_{ij}$ associated with
the particle identification, $M_{ij}$, is shown in the first four rows of Table~\ref{tableMigMatrix}. Note that this matrix
includes efficiency losses associated with cross-feed of the wrong charge combinations and small factors
associated with data control sample corrections to the MC cross-feed efficiencies, therefore the
columns of the table are not expected to sum to 100\%.

\begin{table*}
\caption{Diagonal elements of $\mathbf{\cal E}_{ij}$ excluding particle identification,
 numbers of selected, background 
and signal events determined using Eq.~\ref{eqn1}. The branching fraction results for this
 work are presented in the fifth line where the first error is statistical and the second systematic.
 The PDG average branching fractions/limit~\cite{PDG} are presented in the sixth line.}
\begin{center}
\small
\begin{tabular}{|l|c|c|c|c|} \hline
              & \tautoppp        & \tautokpp        & \tautokpk      & \tautokkk  \\ \hline
 $\epsilon$   & \effppp            & \effkpp            & \effkpk          & \effkkk    \\
 $N^{Data}$   & \Ndatappp          & \Ndatakpp          & \Ndatakpk        & \Ndatakkk    \\
 $N^{Bkg}$    & \Nbkgppp           & \Nbkgkpp           & \Nbkgkpk         & \Nbkgkkk     \\
 $N^{Sig}$    & \Nsigppp           & \Nsigkpp           & \Nsigkpk         & \Nsigkkk     \\ 
 $\BR$ (this work)        & \BRppp             & \BRkpp             & \BRkpk           & \BRkkk       \\ 
 $\BR$ (PDG average)  & \BRpppPDGa~\cite{Briere:2003fr}   & \BRkppPDGa &  \BRkpkPDGa           & \BRkkkPDGa       \\ \hline
\end{tabular}
\label{table1}
\end{center}
\end{table*}

\normalsize

\begin{table}
\caption{Component of the efficiency migration matrix associated with pion-kaon particle identification, $M_{ij}$, in percent. 
         The $\tautophip$ and $\tautophik$ study employs a higher efficiency/lower purity kaon selection algorithm.}
\begin{center}
\begin{tabular}{|l|rrrr|} \hline
 Candidates & \multicolumn{4}{c|}{Decay modes}             \\
           & $\pi\pi\pi$& $K\pi\pi$ & $K\pi K$  & $KKK$  \\ \hline

$\pi\pi\pi$& 97.40      & 22.49     & 4.73      & 1.02  \\

$K\pi\pi$  & 1.42       & 74.87     & 16.43     & 6.38   \\

$K\pi K$   & 0.01       &  0.49     & 59.63     & 25.54  \\

$KKK$      & $ - $      & $ - $     & 0.26      & 50.87  \\ \hline

$\phi\pi$  &            &           & 72.54     & 19.20 \\

$\phi K$   &            &           & 0.83      &  66.06 \\ \hline
\end{tabular}
\label{tableMigMatrix}
\end{center}
\end{table}

Systematic uncertainties are assigned for:  luminosity; cross-section; 
 migration matrix elements, which includes MC statistical and systematic errors associated with
 the efficiency and particle identification; signal-mode modeling; 
 modeling of the EMC and tracking response including scale and resolution uncertainties, 
the sensitivities of the measurements to the modeling of hadronic and electromagnetic showers in the EMC,
and tracking efficiency; modeling of the trigger; and modeling of the backgrounds, including uncertainties on
cross-sections and branching fractions.
These are summarized in Table~\ref{tablesys} along with the total error correlation matrix.
The absolute normalization is a significant source of the correlations.

\begin{table}
\caption{Upper: Systematic uncertainties (\%).
Lower: Correlation matrix from stat. $\oplus$ syst. covariance matrix.}
\begin{center}
\begin{tabular}{|l|c|c|c|c|} \hline
                          & $\pi\pi\pi$      & $K\pi\pi$        & $K\pi K$      & $KKK$  \\ \hline
 $\mathcal{L}$            &   0.9           & 0.9              & 0.9            & 0.9        \\ 
 $\sigma_{\eett}$         &   0.3           & 0.3              & 0.3            & 0.3        \\ 
 $M_{ij}$ and particle ID &   0.4            & 3.0              & 1.9            & 4.9        \\
 signal modeling         &   0.2            & 0.2              & 1.3            & 2.0        \\ 
 EMC and DCH response     &   0.8            & 0.9              & 0.8            & 1.2        \\
 trigger                  &   0.1            & 0.1              & 0.1            & 0.1        \\
 backgrounds              &   0.4            & 0.7              & 0.4            & 5.5        \\ \hline
 Total                    &   1.4            & 3.4              & 2.7            & 7.8       \\ \hline \hline
$\pi\pi\pi$                &                  & 0.544            & 0.390          & 0.031     \\
$K\pi\pi$                  &                  &                  & 0.177          & 0.093     \\   
$K\pi K$                   &                  &                  &                & 0.087     \\ \hline
\end{tabular}
\label{tablesys}
\end{center}
\end{table}

 The branching fraction results of this analysis are presented in Table~\ref{table1} together with 
 the world average values/limit published by the Particle Data Group (PDG)~\cite{PDG}, with which they are consistent
 and significantly more precise.
 In all four channels, the results
 where the lepton-side  has an  electron are consistent with those where it has a muon.
 Our measurement of \BRtauppp is also consistent with a precision $\BR(\tau^- \to \pi^-\pi^-\pi^+[\mathrm{ex.} \omega])$
 measurement~\cite{Schael:2005am} after accounting for the $\omega$.
 Our measurement of  \BRtaukpp is in agreement with~\cite{Barate:1997ma} and
 disagrees by more than two standard deviations from \cite{Briere:2003fr} and \cite{Abbiendi:2004xa}.
 We report a first measurement of \BRtaukkk  in which no resonance structure is assumed and which has a significance
 in excess of 8$\sigma$.

\begin{figure}
\resizebox{\columnwidth}{0.716\columnwidth}{
\includegraphics{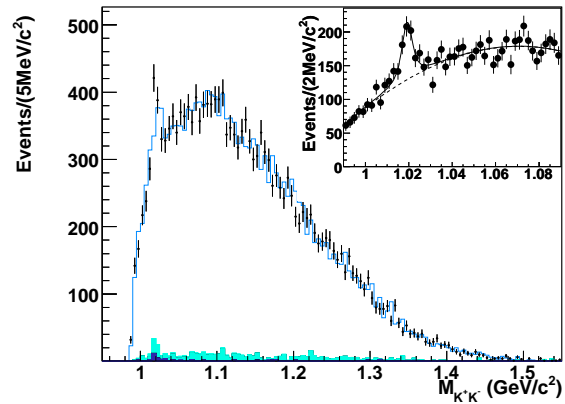}}
\caption{$K^+K^-$ invariant mass in \tautokpk decays.
Data are represented by points with error bars, MC of \tautokpk by the open histogram,
cross-feed from other \tautohhh channels by the light shaded histogram,
and non-$\tau$ backgrounds by the dark shaded histogram.
The inset  shows the background-subtracted data with the
  fit (solid line) and non-resonant component (dashed line).}
\label{figurekpk}
\end{figure}

\begin{figure}
\resizebox{\columnwidth}{0.716\columnwidth}{
\includegraphics{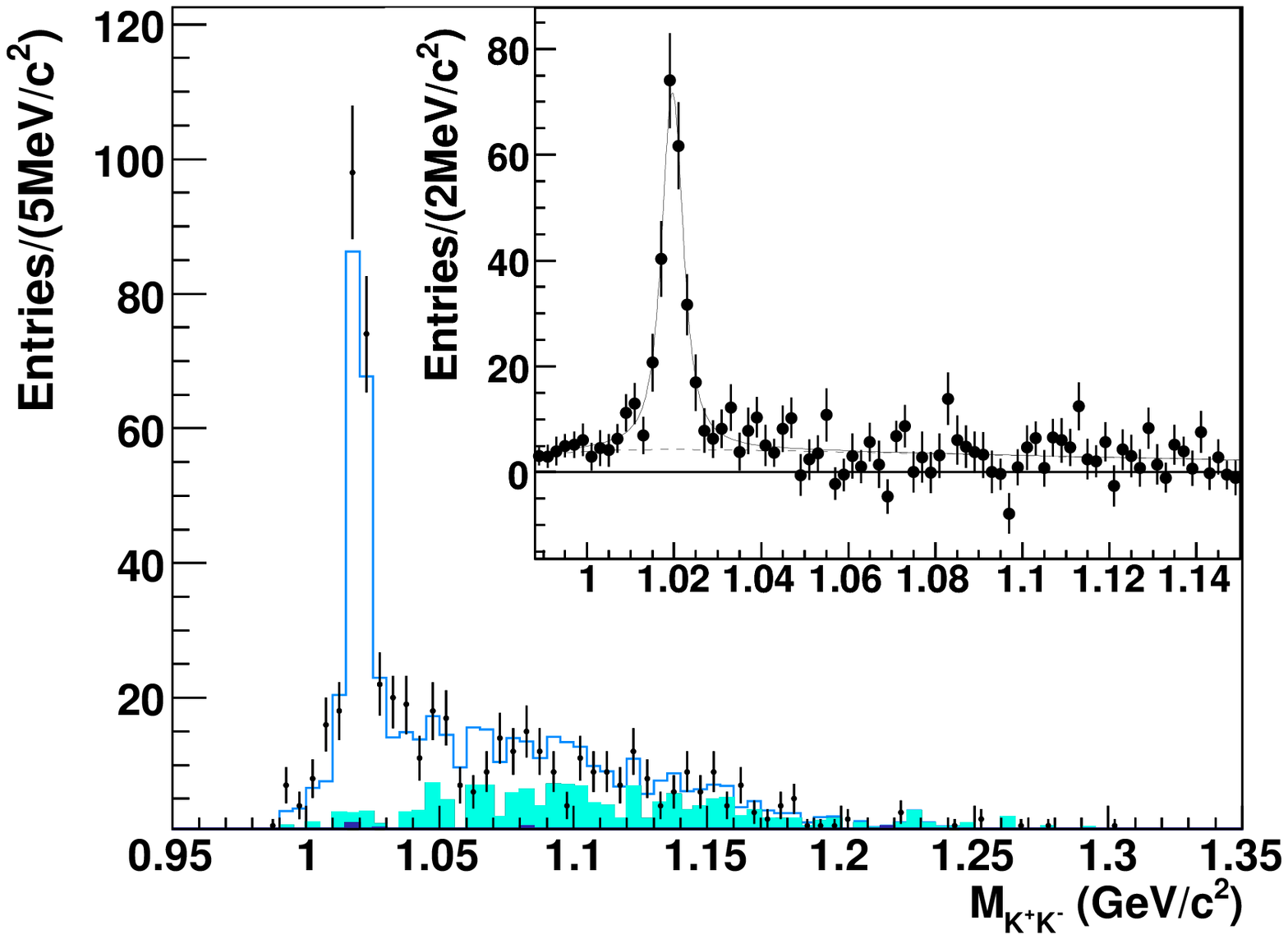}}
\caption{$K^+K^-$ invariant mass in \tautokkk decays with two entries per event.
Data are represented by points with error bars, the open histogram is the MC of \tautophik, the light shaded
 histogram is the cross-feed from the other \tautohhh channels, primarily from \tautokpk,
and non-$\tau$ backgrounds by the dark shaded histogram.
The inset  shows the background-subtracted data with the
  fit (solid line) and non-resonant component (dashed line).}
\label{figurekkk}
\end{figure}

 A $\phi(1020)$ contribution is seen in both the \tautokpk\ and \tautokkk\ decay modes.
 The use of a kaon selection algorithm with higher efficiency, but less purity,
 provides significantly higher signal-to-background for \tautophik.
 The \tautophip (\tautophik)  signal has a 5.7$\sigma$ (9.8$\sigma$)  level of significance.
 The last two rows of Table~\ref{tableMigMatrix} lists the $M_{ij}$ matrix for this higher efficiency selection. 
 The  $K^+K^-$ invariant mass distributions for the \tautokpk (\tautokkk) mode using this  sample is
 shown in Fig.~\ref{figurekpk} (\ref{figurekkk}). Below 1.09\gev (1.15\gev), after background subtraction of
 the non-$K^+K^-$ events, the $K^+K^-$ invariant mass distribution from the  $\tautophip$ ($\tautophik$) decay
 is well described by a Breit-Wigner function convoluted with a Gaussian resolution function for the signal and
 a third-order polynomial (function in~\cite{Albrecht:1990am}) for the background and is used to fit for the number of events.
 A binned maximum likelihood fit  yields 344$\pm$42 (274$\pm$16) $\tautophip$ ($\tautophik$) candidates.
 MC estimates of the subtracted $q\bar{q}$ background events with a $\phi$ contribute a 5.2\% (0.7\%) uncertainty.
 The $\phi\pi^-$ candidate sample has an additional 4.3\%  uncertainty arising from
 potentially peaking \tautophippiz\  background.
 The fit parameterization contributes a 1.0\% (2.0\%) error.
 Accounting for $\BR(\phi\ra K^+K^-)$~\cite{PDG},
 $\BRtauphip=$ $\BRphip$ and $\BRtauphik=$ $\BRphik$ with a correlation of -0.07.
 From the fit we find no evidence for  $\tautokkk$ 
 without a $\phi$ and set a first upper limit on $\BRtaukkkexphi$ $\BRkkkexphi$ at 90\% CL.

 This is the first measurement of $\BRtauphip$. It is consistent with
 a CLEO limit~\cite{Avery:1996de} and 
 larger than the $(1.20\pm0.28)\times 10^{-5}$ value predicted by a vector meson dominance model~\cite{LopezCastro:1996xh}.
 Our $\BRtauphik$ measurement is consistent with a recent Belle result~\cite{Abe:2006uk}.
 Recent calculations using a meson dominance model~\cite{FloresTlalpa:2007bt} agree with our
 $\BRtauphip$ and $\BRtauphik$ measurements and with 
 the ratio $\frac{\BRtauphip}{\BRtauphik} = 0.99 \pm 0.21$.
 Our measurements of $\BRtaukpp$ and $\BRtaukkk$,  when combined with improved measurements of the other strange decays, 
 will constrain the CKM element $|V_{us}|$ better than unitarity bounds~\cite{Maltman:2006ic,Hill:2006bq}.

We are grateful for the excellent luminosity and machine conditions
provided by our \pep2\ colleagues, 
and for the substantial dedicated effort from
the computing organizations that support \babar.
The collaborating institutions wish to thank 
SLAC for its support and kind hospitality. 
This work is supported by
DOE
and NSF (USA),
NSERC (Canada),
IHEP (China),
CEA and
CNRS-IN2P3
(France),
BMBF and DFG
(Germany),
INFN (Italy),
FOM (The Netherlands),
NFR (Norway),
MIST (Russia),
MEC (Spain), and
PPARC (United Kingdom). 
Individuals have received support from the
Marie Curie EIF (European Union) and
the A.~P.~Sloan Foundation.

\end{document}